\documentclass[a4paper,12pt]{iopart}

\usepackage{amssymb}
\usepackage{lineno,hyperref}
\usepackage{epsfig}
\usepackage{graphicx}
\usepackage[usenames, dvipsnames]{color}

\begin{document}

\title[]{A new scheme for short baseline
electron antineutrino disappearance study}

\author{Jae Won Shin$^1$, Myung-Ki Cheoun$^1$, Toshitaka Kajino$^2$ and Takehito Hayakawa$^3$}

\address{$^1$ Department of Physics and Origin of Matter and Evolution of Galaxies (OMEG) Institute,
Soongsil University, Seoul 156-743, Korea}
\address{$^2$ Division of Theoretical Astronomy,
National Astronomical Observatory
of Japan, Mitaka, Tokyo 181-8588, Japan \\
and Department of Astronomy, Graduate School of Science, University of
Tokyo, Hongo, Bunkyo-ku, Tokyo 113-0033, Japan}
\address{$^3$ Quantum Beam Science Directorate (QUBS),
Japan Atomic Energy Agency (JAEA), 2-4 Shirane, Shirakata, Tokai-mura, Naka-gun, Ibaraki 319-1195, Japan}
\ead{cheoun@ssu.ac.kr (Corresponding Author)}

\vspace{10pt}
\begin{indented}
\item[]1 Aug. 2017
\end{indented}

\begin{abstract}
A new scheme for the short baseline
electron antineutrino
(${\bar{\nu}}_{e}$)
disappearance study
is investigated.
We propose
to use an intense neutron emitter,
$^{252}$Cf,
which produces $^{8}$Li isotope through the $^{7}$Li(n,$\gamma$)$^{8}$Li reaction;
$^{8}$Li is a ${\bar{\nu}}_{e}$ emitter via $\beta^{-}$ decay.
Because this ${\bar{\nu}}_{e}$ source needs neither accelerator nor reactor facilities,
the ${\bar{\nu}}_{e}$
can be placed on any neutrino detectors
as closely as possible.
This short baseline circumstance
with a suitable detector
enables us to study
the existence of possible sterile neutrinos,
in particular, on 1 eV mass scale.
Also, complementary comparison studies
among different neutrino detectors
can become feasible
by using
${\bar{\nu}}_{e}$ from the $^{8}$Li source.
As an example,
applications to
hemisphere and cylinder shape scintillator detectors
are performed in detail with the expectation signal modification by the sterile neutrino.
Sensitivities to mass and mixing angles of sterile neutrinos
are also presented for comparison with those of other neutrino experiments.
\end{abstract}


\vspace{2pc}

\submitto{\jpg}

\maketitle

\normalsize

One of open issues in the particle and neutrino physics
is the existence of hypothetical fourth neutrino
which may be mixed with active neutrinos \cite{Review_nu_1}.
This fourth neutrino, so-called sterile neutrinos
(${\nu}_{s}$), is claimed to
play important roles of explaining
some anomalies reported in
LSND \cite{LSND_1}, MiniBoone \cite{MiniB_1},
reactor experiments \cite{reacAnt}, and gallium experiments \cite{GaAnomaly}.
A few experiments with 
compact neutrino detectors and reactor facilities \cite{danss, neu_4, nuci, panda, prospect_1, stereo} or 
antineutrino sources from radioactive isotopes \cite{Ce144s1,Ce144s2,Ce144s3} or 
an accelerator-based IsoDAR (isotope decay-at-rest) \cite{isoDar1, sterileNu1, annu8Li1} 
are proposed 
to search the existence of ${\nu}_{s}$. 

By using compact neutrino detectors such as
DANSS \cite{danss},
NEUTRINO4 \cite{neu_4},
NUCIFER \cite{nuci},
PANDA \cite{panda},
PROSPECT \cite{prospect_1}, 
and STEREO \cite{stereo}, 
experiments have been planned to measure reactor neutrinos
at a distance of several meters. 
KamLAND (CeLAND) \cite{Ce144s2} and
Borexino
(SOX) \cite{Ce144s3} plan to perform experiments
by using
antineutrino generators of unstable isotopes $^{144}$Ce-$^{144}$Pr
of a radioactivity of 100 kCi,
where neutrino energies are lower than 3 MeV.
As another type of antineutrino source,
$^{8}$Li was suggested by using
an accelerator-based IsoDAR concept \cite{sterileNu1, annu8Li1}.
Electron antineutrinos
(${\bar{\nu}}_{e}$)
were assumed to be emitted from
$^{8}$Li through $\beta^{-}$ decay
with
energies of up to $\sim$13 MeV.
Because these energies are higher than those from
$^{144}$Ce-$^{144}$Pr, $^{8}$Li can be used for study of
${\bar{\nu}}_{e}$ spectrum distortion
in the energy region of 5 MeV $< E_{\bar{\nu}} <$ 7 MeV,
where
some distortions or anomalies were reported by
reactor antineutrino experiments
(Daya Bay \cite{ReactBump_Day}, Double Chooz \cite{ReactBump_DOU1}
and RENO \cite{ReactBump_RENO1, ReactBump_RENO2}).


In this letter, 
as a new $^{8}$Li generator,
we propose a fissionable isotope of $^{252}$Cf radioactive isotope-based
${\bar{\nu}}_{e}$ production scheme.
In particular, our $^{8}$Li generator 
can be placed on
any neutrino detectors
such as Borexino \cite{Borex}, JUNO \cite{JunoR}, KamLAND \cite{Kam}, LENA \cite{LENA_0} and SNO+ \cite{SNO}, etc., 
because one does not need any accelerator or reactor systems.
As an intense neutron emitter,
$^{252}$Cf is used
for productions of $^{8}$Li.
$^{252}$Cf with a half-life of 2.64 yr
emits neutrons with an average energy of 2 MeV
via spontaneous fissions.
The neutron emission rate
of 1 gram of $^{252}$Cf is
2.34 $\times$ 10$^{12}$ neutrons
per second (n/s).
For energy distribution of neutrons
from $^{252}$Cf we take Watt fission spectrum \cite{watt_1, watt_2, x5} in this work.


\begin{figure}[tbp] 
\centering
\includegraphics[scale=0.4]{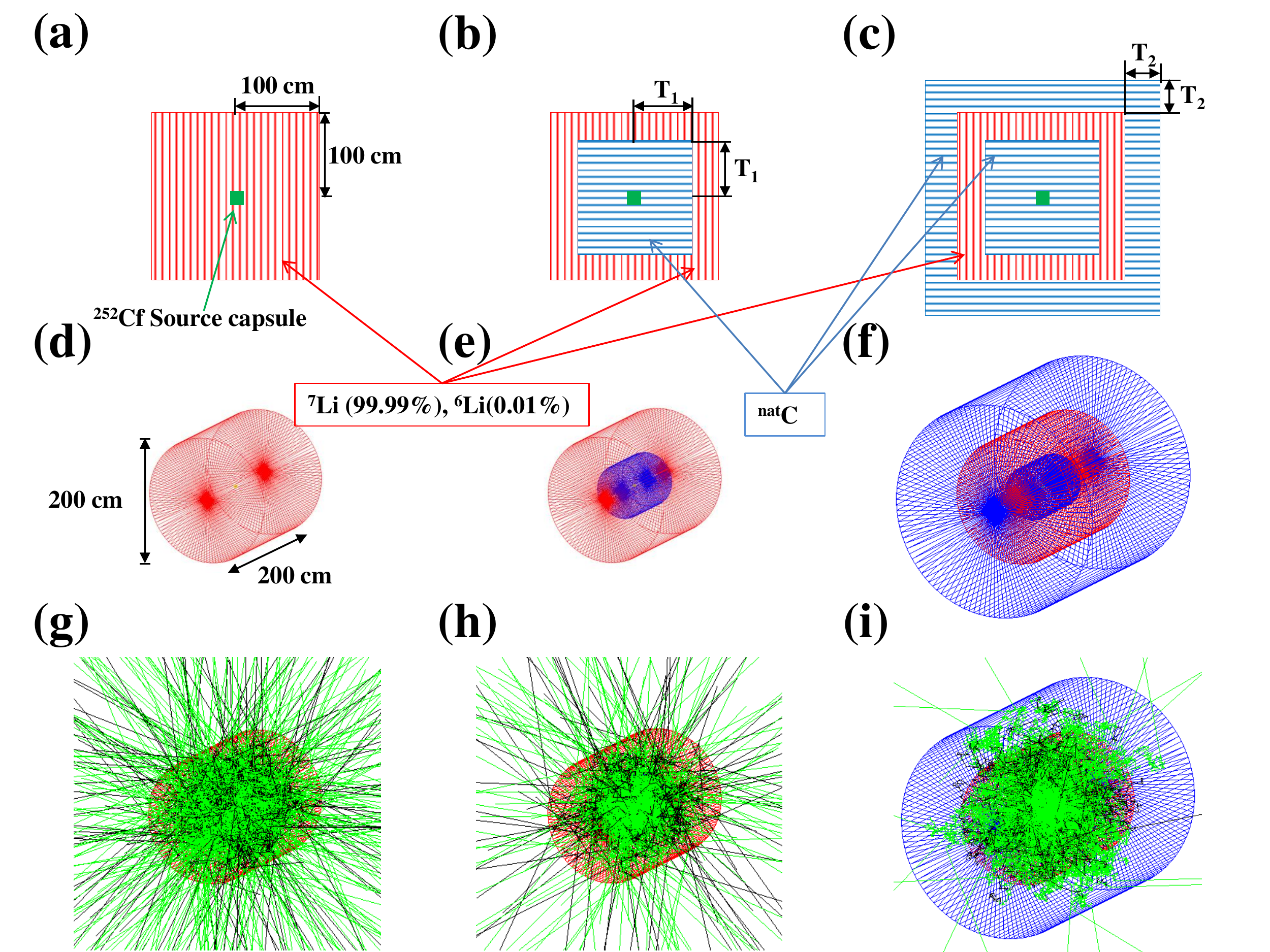}
\caption{(Color online)
Schematic horizontal cross section views (a, b, c) of three different cylindrical $^{8}$Li generator types (d, e, f),
referred as setup I, II and III, respectively.
Simulation snap shots by OpenGL pictures for the $^{8}$Li generators
with $^{252}$Cf source are shown at the lowest panel (g, h, i). Green and black lines
stand for neutrons and $\gamma$, respectively.
}
\label{fig1}
\end{figure}

We consider three different types of geometrical setups
for productions of $^{8}$Li as shown in Fig.~\ref{fig1}.
Figure \ref{fig1} (a, d) (setup I) shows a simple setup case
with 99.99\% enhanced $^{7}$Li convertor
and $^{252}$Cf surrounded by the Li convertor.\footnote{\protect
A $^{252}$Cf source is enclosed with the stainless steel case,
but the presence of the case does not affect present work.
}
The Li convertor has a cylindrical shape of a radius of 100 cm
and a length of 200 cm based in Ref.~\cite{sterileNu1}.
In Fig.~\ref{fig1} (b, e) (setup II),
we add carbon (graphite) material
as an neutron reflector to the setup I.
The carbon is preferred as the neutron reflector
due to its low absorption cross section
and high elastic scattering cross section.
This concept is known as
the Adiabatic Resonance Crossing (ARC)
\cite{ARC_0}
for the transmutation of long-lived radio isotopes in the nuclear waste
and
the medical radioisotope production
\cite{ARC_1, ARC_2, ARC_3}.
With the carbon reflector,
we can increase production yield of $^{8}$Li
even though the amount of $^{7}$Li is less than that of the setup I.
The setup III in Fig.~\ref{fig1} (c, f)
has two carbon reflectors.
One is surrounded by the Li convertor
and another wraps the Li convertor.

The production rate of $^{8}$Li
by neutrons from $^{252}$Cf 
is estimated by using GEANT4 code (v. 10.1) \cite{g4n1, g4n2}.
High precision models with G4Neutron Data Library (G4NDL) 4.5
are used in the present work.
The data in G4NDL 4.5 come largely from
the Evaluated Nuclear Data File (ENDF/B-VII) library \cite{endf}.
Simulation snap shots
for the $^{8}$Li generator setup I, II, and III are shown in Fig.~\ref{fig1} (g, h, i), respectively.\footnote{\protect
For simplicity,
simulations have been performed
with 500 neutrons generation from $^{252}$Cf
where the neutron has a Watt fission spectrum in
Refs.~\cite{watt_1, watt_2, x5}.
}
For the setup I,
neutrons and $\gamma$ rays easily escape from the neutrino source,
while the numbers of neutrons and $\gamma$ escaping from
the setup II and III
are drastically reduced by the carbon reflectors.

First,
we calculate the production yield of $^{8}$Li
for the case of setup I depicted in Fig.~\ref{fig1} (g),
and obtain the yield of 0.0045 $^{8}$Li per neutron ($^{8}$Li/n). The production yields of $^{8}$Li
for the setup II and III
are sensitive to the thickness of carbon reflector,
T$_1$ and T$_2$, in Fig.~\ref{fig1}.
As the thickness of the reflector increases,
numbers of the collisions
between the neutron and
the carbon also increase.
Thus, probabilities for the neutron capture by $^{7}$Li
can also increases.
If the T$_1$ is too thick, however,
it becomes hard for the scattered neutron
to escape from the inner carbon, namely, these neutrons are rarely absorbed to $^{7}$Li.

\begin{figure}[tbp]
\centering
\includegraphics[scale=0.25]{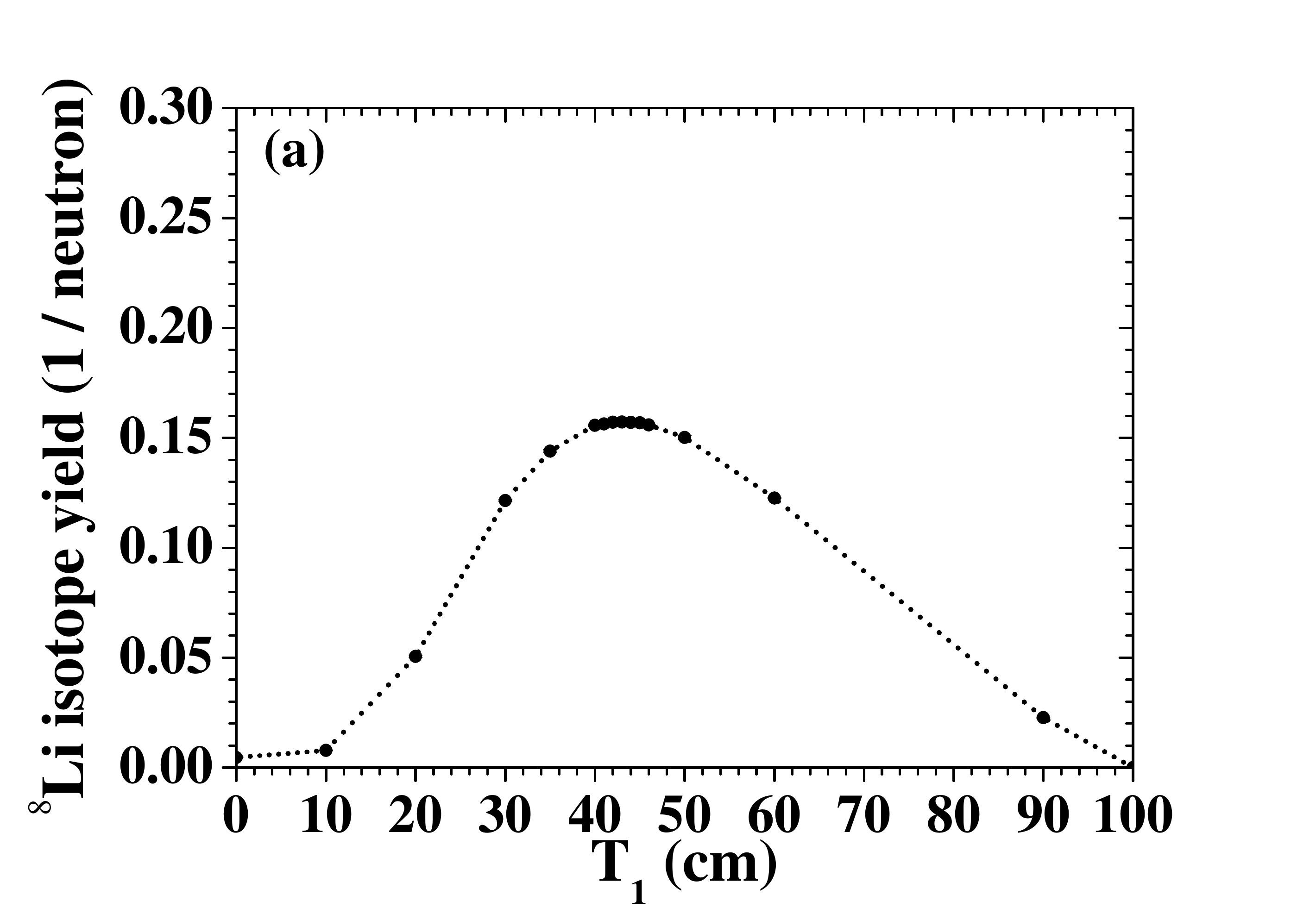}
\includegraphics[scale=0.25]{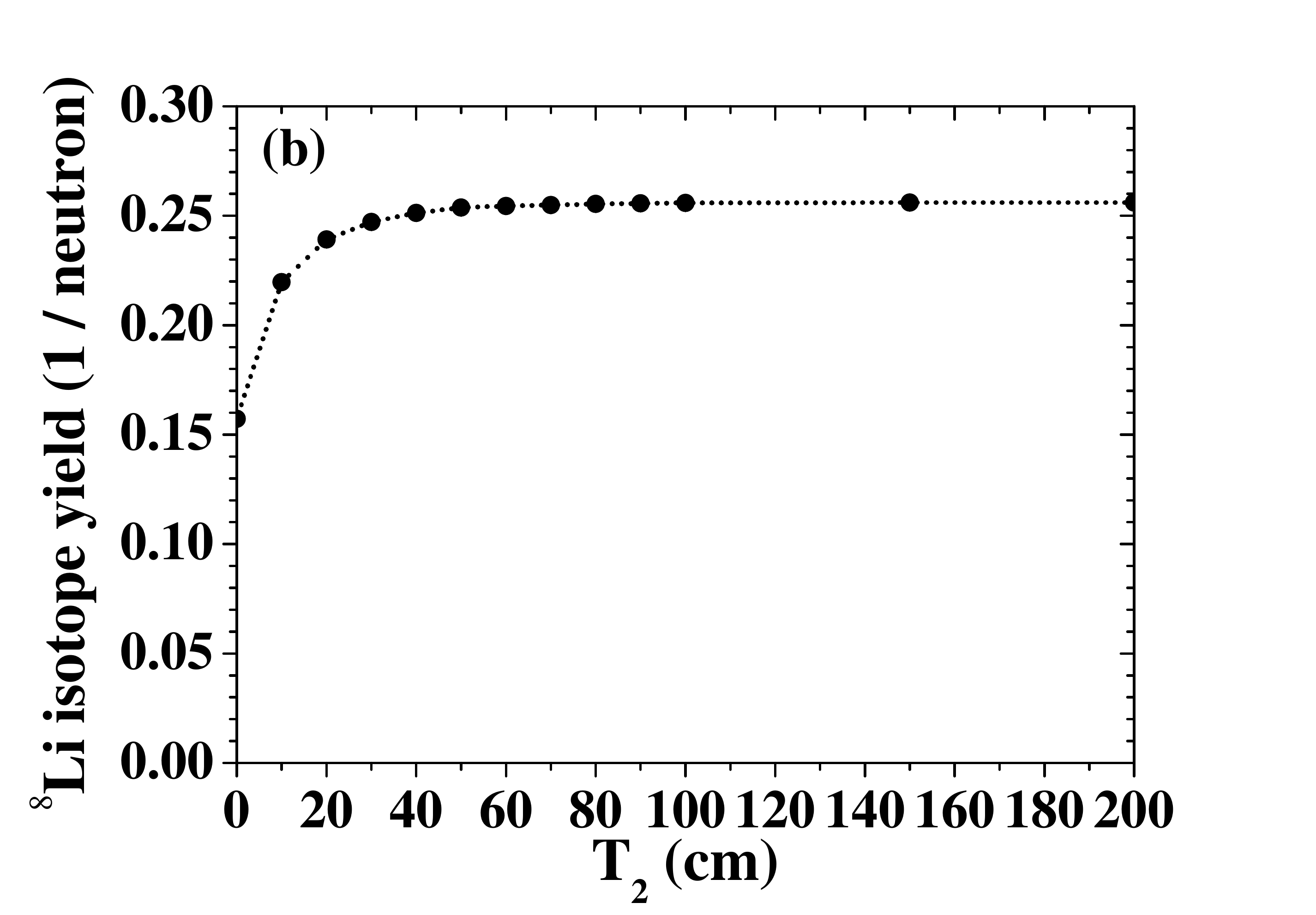}
\caption{(Color online)
(a) $^{8}$Li isotope yields for setup II with respect to T$_{1}$.
(b) $^{8}$Li isotope yields for setup III with respect to T$_{2}$, where T$_{1}$ is set to be 43 cm which is a optimal thickness in (a).
}
\label{set1gra}
\end{figure}

Figure \ref{set1gra} (a) shows
the dependence of the $^{8}$Li yields on the
thickness of carbon reflector (T$_{1}$) for setup-II.
Yields of $^{8}$Li increase
when T$_{1}$ increases up to 43 cm.
Maximum yield of $^{8}$Li is
0.157 $^{8}$Li/n at T$_{1}$ = 43 cm,
where the yield is larger than those for setup-I
about 35 times.
As the T$_{1}$ increases more than 43 cm,
yields of $^{8}$Li decrease.
In Fig.~\ref{fig1} (h),
un-captured neutrons by $^{7}$Li
still can escape from the geometry.
To further increase the yields of $^{8}$Li,
additional carbon reflector
with the thickness T$_{2}$
is considered to be located out of the Li convertor.

The production yield for $^{8}$Li
with respect to T$_{2}$
is plotted in Fig.~\ref{set1gra} (b).
Figure~\ref{set1gra} (a) shows that
the optimal thickness of T$_{1}$ is 43 cm,
and thus the same thickness
is chosen for setup-III.
As the T$_{2}$ increases,
$^{8}$Li yields also increase up to 0.256 $^{8}$Li/n.
With the T$_{2}$ larger than 50 cm,
yields for $^{8}$Li are almost saturated.
Maximum yields of $^{8}$Li for setup-III
are larger than those for setup-II and setup-I
by factors
$\sim$1.6 and $\sim$56.9,
respectively.

We obtained a result that the yields of $^{8}$Li increase
to the maximum value at T$_1$ = 43 and T$_{2}$ = 100 cm of the setup III,
where the production yield is 0.256 $^{8}$Li/n.
$^{8}$Li with a half-life of 0.838 s emits
${\bar{\nu}}_{e}$ through $\beta^{-}$ decay.
The electron anti-neutrinos from $^{8}$Li
have continuous energy distribution,
which is evaluated by using
``G4RadioactiveDecay" \cite{G4RDM_0, G4RDM_1} class
based on the Evaluated Nuclear Structure Data File (ENSDF) \cite{G4ensdf}.

\begin{figure}[tbp]
\centering
\includegraphics[scale=0.3]{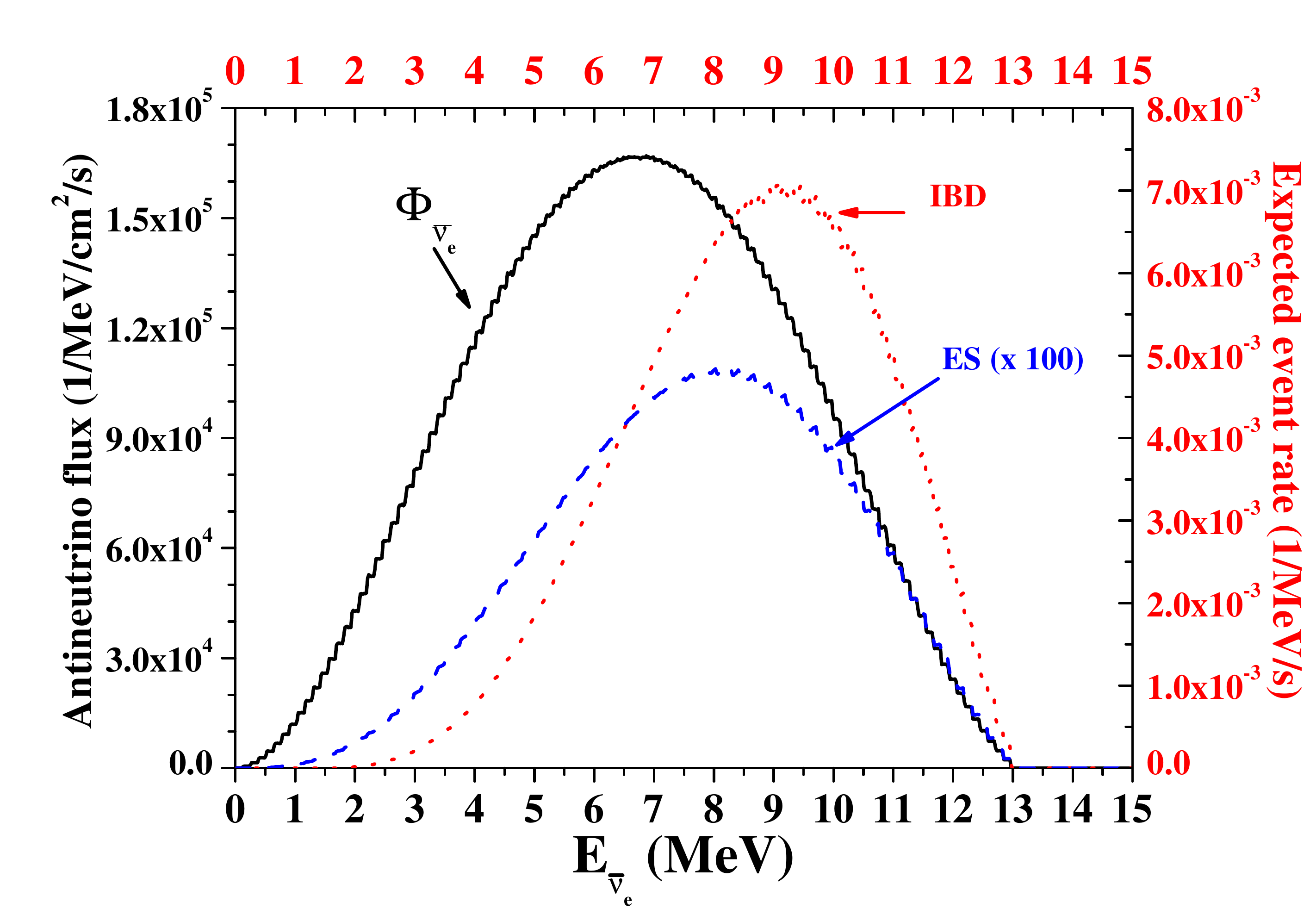}
\caption{(Color online)
${\bar{\nu}}_{e}$
flux from $^{8}$Li source
and expected event rates for
ES and IBD. Here we assumed 1 g of $^{252}$Cf in the setup III and detector at 2 m from the neutrino source.
The black solid line is the electron-antineutrino energy spectrum.
The red dotted line and the blue dashed line are
the event rates for IBD and ES, respectively.
For ES,
we multiply 100 to the rate.
}
\label{fig5}
\end{figure}

Electron antineutrinos
can be measured with
two different neutrino reactions.
One is 
${\bar{\nu}}_{e}$ elastic scattering on electrons (ES),
and another is inverse beta decay (IBD) reaction,
${\bar{\nu}}_{e} + p \to e^{+} + n$.
In ES,
${\bar{\nu}}_{e}$ can be indirectly measured
by the scattered e$^{-}$ by liquid-scintillator (LS).
Expected event rate for ES can be obtained from Refs.~\cite{LENA_ER, nu_eElas}.
The event rate ($R^{IBD}_{{\bar{\nu}}_{e}}$) for IBD
is written as
\begin{eqnarray}
R^{IBD}_{{\bar{\nu}}_{e}}
= n_{p} \int^{E_{max}}_{E_{th}} dE_{\bar{\nu}} \Phi_{{\bar{\nu}}_{e}} (E_{\bar{\nu}}) {\rm{P}}_{\nu \bar{\nu}}(E_{\bar{\nu}}) \sigma^{IBD}_{{\bar{\nu}}_{e}}(E_{{\bar{\nu}}_{e}})~,
\label{eq:IBD_rate}
\end{eqnarray}
where $n_{p}$ is the number of target protons within a fiducial volume of the detector,
$\Phi_{{\bar{\nu}}_{e}} (E_{\bar{\nu}})$ is the ${\bar{\nu}}_{e}$ flux from $^{8}$Li,
${\rm{P}}_{\nu \bar{\nu}}(E_{\bar{\nu}})$ is the energy dependent ${\bar{\nu}}_{e}$ survival probability.
The energy dependent cross section of IBD
is approximately taken by \cite{ibdCS}
\begin{eqnarray}
\sigma^{IBD}_{{\bar{\nu}}_{e}}(E_{{\bar{\nu}}_{e}})
  &\approx& p_{e} E_{e}
  E_{{\bar{\nu}}_{e}}^{-0.07056+0.02018{\rm{ln}}E_{{\bar{\nu}}_{e}} - 0.001953 {\rm{ln}}^{3} E_{{\bar{\nu}}_{e}}}
  \times 10^{-43} [\rm{cm}^{2}],
\label{eq:IBDcs}
\end{eqnarray}
where $p_{e}$, $E_{e}$, and $E_{{\bar{\nu}}_{e}}$ are
the positron momentum, total energy of
the positron,
and energy of ${\bar{\nu}}_{e}$ in MeV, respectively.\footnote{\protect 
This cross section agrees within few per-mille
with the full calculation including
the radiative corrections and the final-state interactions in IBD.}

In Fig.~\ref{fig5},
energy distribution of ${\bar{\nu}}_{e}$ from $^{8}$Li
and expected event rates of ES and
IBD are presented,
where $^{8}$Li is assumed to be produced by
the setup III.
For the calculation of neutrino oscillation,
we use the ${\rm{P}}_{\nu {\bar \nu}}(E_{\bar{\nu}})$ ($\equiv$P$_{3}$)
given by \cite{PeeRef1}
\begin{eqnarray}
P_{3}
= 1 - {\rm{sin}}^{2} 2 \theta_{13} S_{23} - c^{4}_{13}{\rm{sin}}^{2} 2 \theta_{12} S_{12},
\label{eq:Pee_1}
\end{eqnarray}
where
$S_{23} = {\rm{sin}}^{2}(\Delta m^{2}_{32} L / 4 E)$ and
$S_{12} = {\rm{sin}}^{2}(\Delta m^{2}_{21} L / 4 E)$.
Neutrino oscillation parameters
are taken from a global fit from Ref.~\cite{nu3active}.
For comparison,
we chose the number of electrons ($n_{e}$) in LS for ES
as the same as the numbers of protons ($n_{p}$) in LS for IBD.
The reaction rates of ES turn out to be much smaller than those of IBD.
Therefore, herefrom we only consider IBD for the following neutrino disappearance study.

The IBD reaction offers two signals in neutrino detections;
one is a prompt signal due to annihilation of a positron,
and another is a delayed signal of 2.2 MeV $\gamma$ ray
via a neutron capture, which provides almost unambiguous antineutrino event detection.
The two distinct detections
give an efficient rejection
of other possible backgrounds.

Note that various unstable isotopes
which emit antineutrinos
can be produced.
In the Li convertor,
$^{3}$H, $^{6}$He, and $^{10}$Be are produced
as well as $^{8}$Li.
However,
$^{3}$H decays with a half-life of 12.3 y,
and the production yields of $^{6}$He and $^{10}$Be
are much lower than that of $^{8}$Li
by factors of 10$^{4}$ and 10$^{7}$, respectively.
$^{10}$Be, $^{12}$B, and $^{14}$C
are produced in the carbon reflectors.
Due to low yields of $^{10}$Be and $^{12}$B
and a long half-life of $^{14}$C (5.7 $\times$ 10$^{3}$ y),
their contributions are marginal for the IBD neutrino
detections.

\begin{figure}[tbp]
\centering
\includegraphics[scale=0.25]{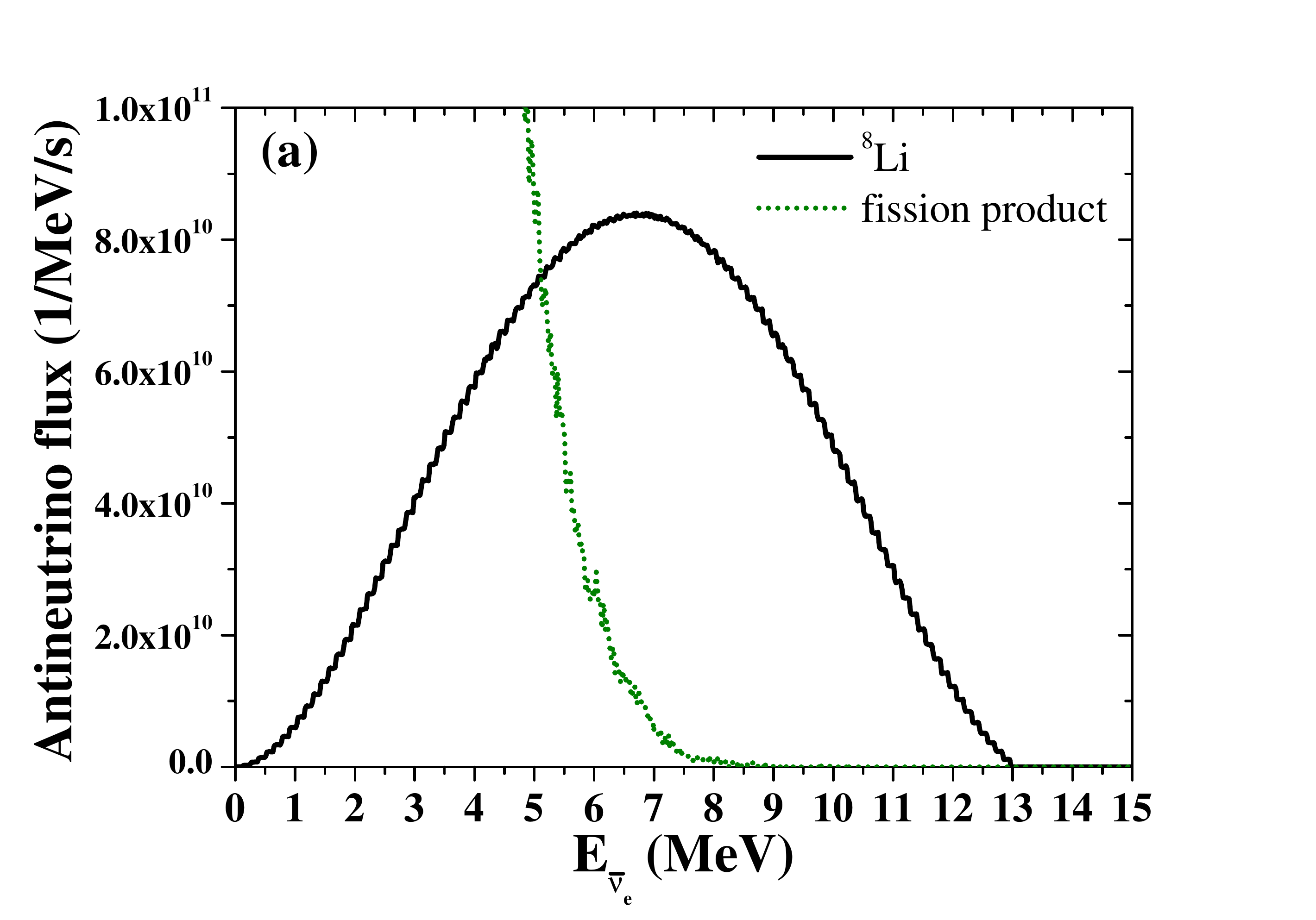}
\includegraphics[scale=0.25]{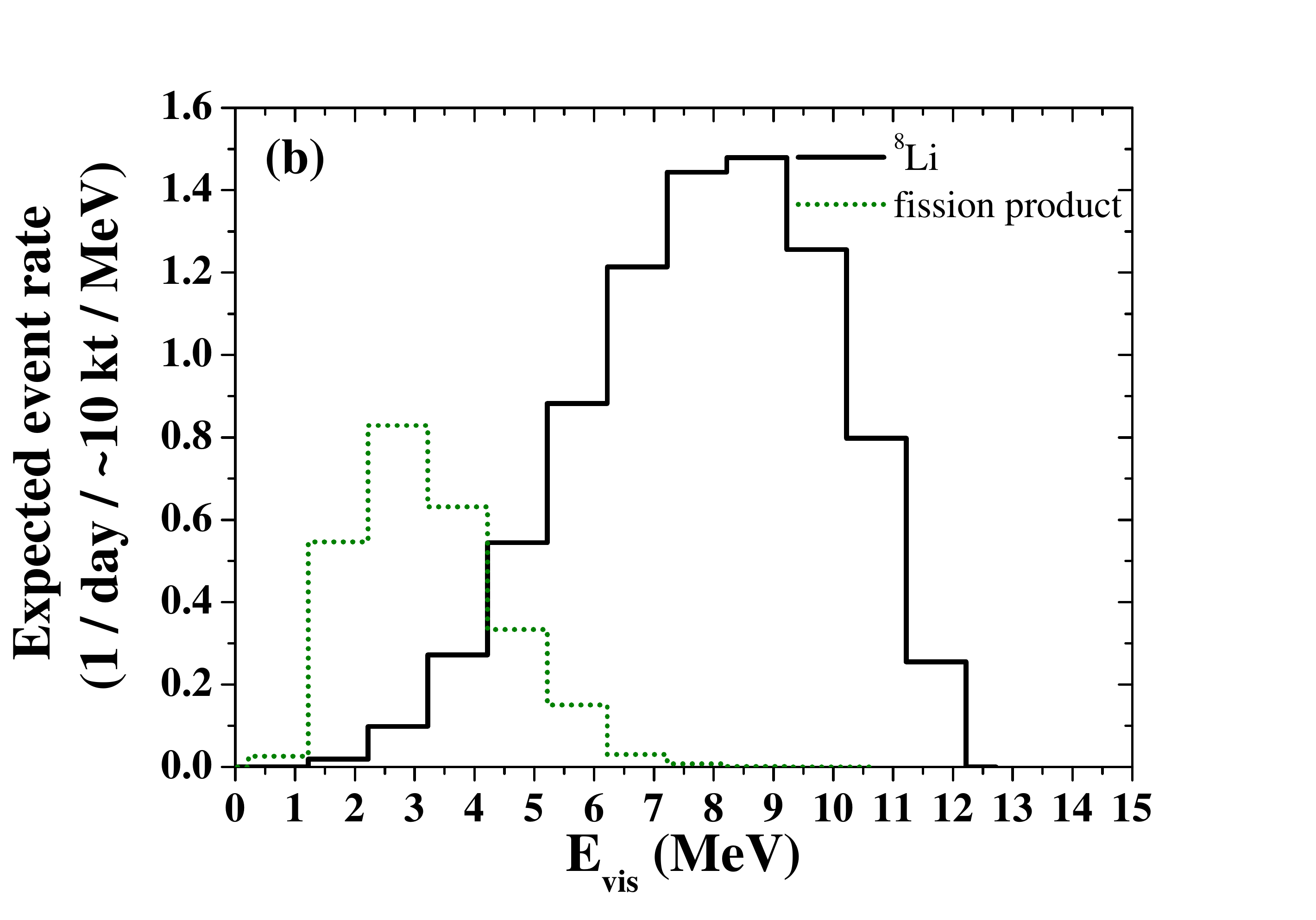}
\caption{(Color online)
${\bar{\nu}}_{e}$ flux (a), which is total flux of that in figure~\ref{fig5}, and expected event rates (b) for
$^{8}$Li and fission product of $^{252}$Cf.
The black solid lines denote the electron-antineutrino from $^{8}$Li
and the green dotted lines represent those from fission product of $^{252}$Cf.
}
\label{figFISS}
\end{figure}

There are background neutrinos such as
neutrinos from fission product of $^{252}$Cf ($\nu_{f}$) and 
geo-neutrinos ($\nu_{geo.}$).
To see the effect of the $\nu_{f}$,
we estimate flux and event rate for $\nu_{f}$ by using
ENDF/B-VII and ENSDF data,
and the results are compared with those from $^{8}$Li ($\nu_{^{8}Li}$) 
in Fig. \ref{figFISS}.
Figure \ref{figFISS} (a) shows that
the $\nu_{f}$ are dominant in low energy regions.
However, for E$_{{\bar{\nu}}_{e}} >$ 7 MeV (corresponding to E$_{vis} >$ 6.22 MeV),
contributions of the $\nu_{f}$ are negligible compare to
the $\nu_{^{8}Li}$.
Total expected event rate of $\nu_{f}$ for E$_{{\bar{\nu}}_{e}} >$ 7 MeV 
in Fig. \ref{figFISS} (b) 
is smaller than those from $^{8}$Li
by three orders of magnitudes.
With the neutrino energy cut of 7 MeV,
we can remove the effect of $\nu_{f}$
in our work.
KamLAND \cite{geoKam_0} and Borexino \cite{geoBorex_0, geoBorex_1} have measured
a rate for $\nu_{geo.}$ ($\sim$ a few events/(100 ton $\cdot$ yr))
due to the decay of U or Th in the Earth.
Contributions of the $\nu_{geo.}$ is negligible compare to those
of ${\bar{\nu}}_{e}$ from $^{8}$Li. 

Electron-antineutrino survival probabilities in
Eq.~(\ref{eq:IBD_rate})
by the 3+1 and 3+2 scenarios can be written as \cite{sterileNu1}
\begin{eqnarray}
P_{\rm{3+1}}
= 1 - 4|U_{e4}|^{2}(1-|U_{e4}|^{2}){\rm{sin}}^{2}(\Delta m^{2}_{41} \frac{L}{4E}),
\label{eq:P_31}
\end{eqnarray}
\begin{eqnarray}
P_{\rm{3+2}} =
1 &-& 4[ (1-|U_{e4}|^{2}-|U_{e5}|^{2}) \nonumber \\
&\times& (|U_{e4}|^{2}{\rm{sin}}^{2}(\Delta m^{2}_{41} \frac{L}{4E})
+ |U_{e5}|^{2} {\rm{sin}}^{2}(\Delta m^{2}_{51} \frac{L}{4E})) \nonumber \\
&+& |U_{e4}|^{2} |U_{e5}|^{2} {\rm{sin}}^{2}(\Delta m^{2}_{54} \frac{L}{4E})],
\label{eq:P_32}
\end{eqnarray}
where relevant parameters
are taken from the best-fit points for the 3+1 and 3+2 scenarios
from the reactor antineutrino data
at Table 1 in Ref.~\cite{sterileNu2}.


\begin{figure}[tbp]
\centering
\includegraphics[scale=0.25]{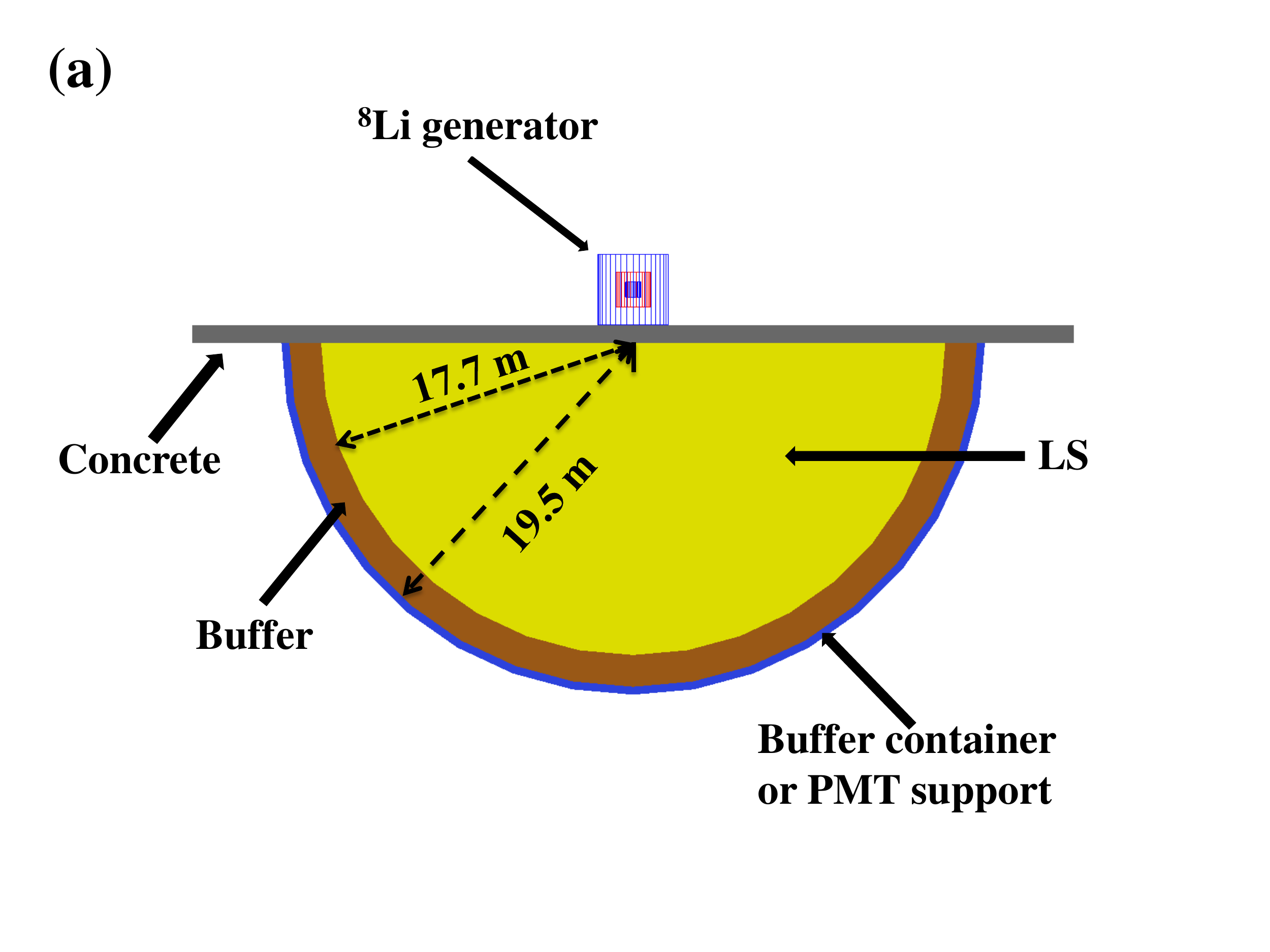}
\includegraphics[scale=0.25]{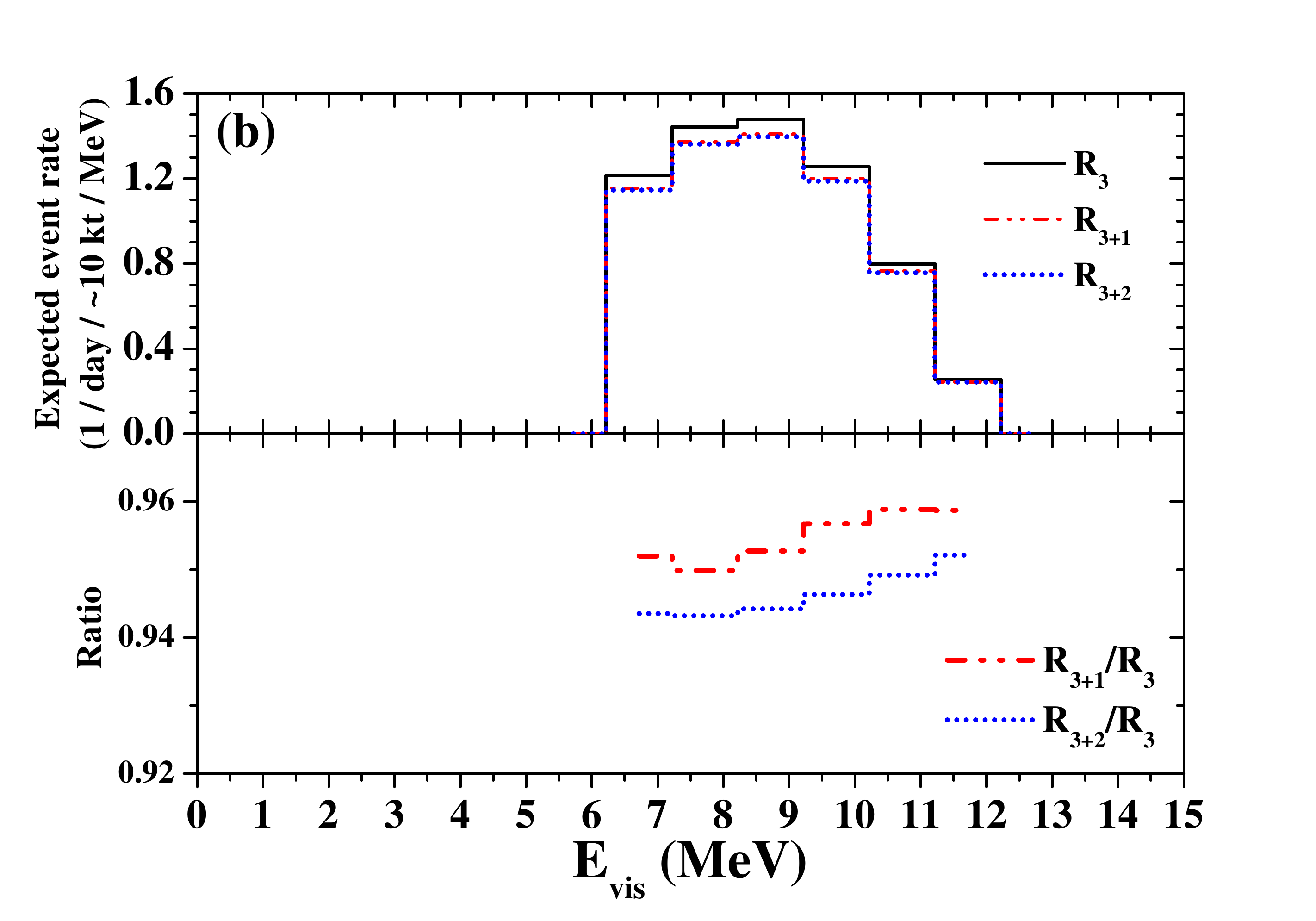}
\caption{(Color online)
(a) OpenGL picture showing simulation geometry for hemisphere type detector with $^{8}$Li generator.
(b) Expected event rates,
R$_{3}$, R$_{3+1}$ and R$_{3+2}$ by P$_{3}$, P$_{3+1}$ and P$_{3+2}$ models,
and their ratios with respect to E$_{vis}$.
Here 6 $\times$ 10$^{11}$ ${\bar{\nu}}_{e}$/s
from the setup III source
with T$_{1}$ = 43 cm and T$_{2}$ = 100 cm by 1 g of $^{252}$Cf
is applied for the calculation.}
\label{figIBD_1}
\end{figure}
%
%
The visible energy (E$_{vis}$) of the prompt signal
due to a
positron ($e^{+}$)
is strongly correlated with
the energy of ${\bar{\nu}}_{e}$ (E$_{{\bar{\nu}}_{e}}$),
E$_{{\bar{\nu}}_{e}}$ $\simeq$ E$_{vis}$ + 0.78 MeV,
by which ${\bar{\nu}}_{e}$ energy spectrum can be reconstructed
using E$_{vis}$.
The spectral shape of E$_{vis}$ would give a valuable chance
to check the existence of the fourth neutrino.
To see the effect of ${\nu}_{s}$
using the shape analysis like Double Chooz, Daya Bay, and RENO experiments,
event rates are estimated for two different types of LS detectors
based on the JUNO \cite{JunoR} and the LENA \cite{LENA_0}
with respect to E$_{vis}$ in the following.
%

\begin{figure}[tbp]
\centering
\includegraphics[scale=0.25]{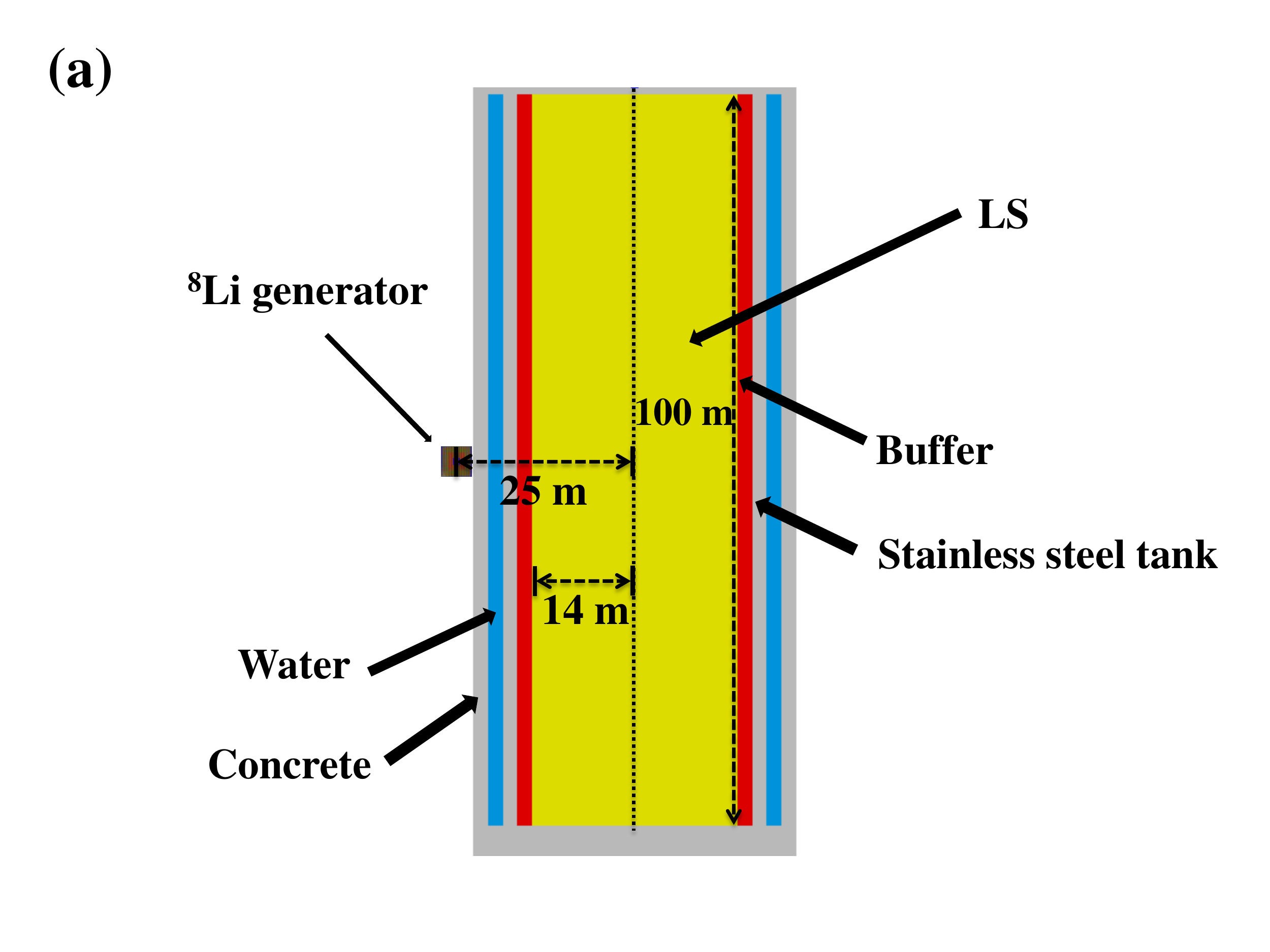}
\includegraphics[scale=0.25]{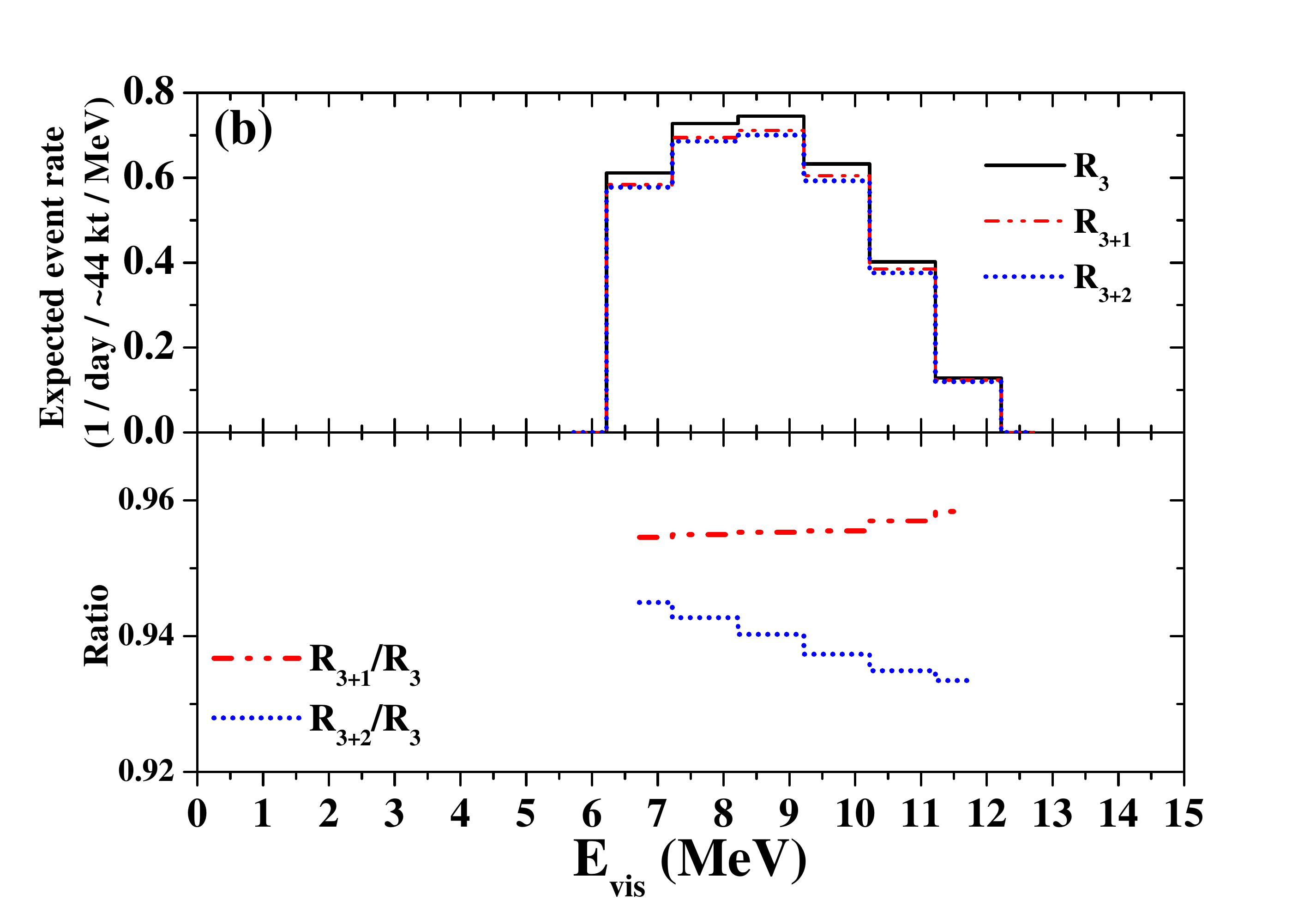}
\caption{(Color online)
%
Same as in Fig.~\ref{figIBD_1} except that the detector has a cylinder shape.
}
\label{figIBD_2}
\end{figure}

%
First, a hemisphere shape scintillator detector 
based on the JUNO \cite{JunoR} 
is considered. 
The present proposed $^{8}$Li generator
requires the space of T$_{1}$ = 43 cm and T$_{2}$ = 100 cm
in a detector.
Thus,
we consider a hemisphere shape of LS detectors with
a radius of 17.7 m
in Fig.~\ref{figIBD_1}(a) where the $n_{p}$
in the detector is reduced to
0.725 $\times$ 10$^{33}$ ($\sim$10 kt) \cite{JunoR}. 
The expected event rate is obtained within the hemisphere shape
of LS detectors
with the cylindrical setup III $^{8}$Li generator.
The evaluated event rates with P$_{3}$, P$_{3+1}$ and P$_{3+2}$,
and their ratios are plotted in Fig.~\ref{figIBD_1} (b).
The ratios of R$_{3+1}$/R$_{3}$ and R$_{3+2}$/R$_{3}$
turns out to be $\sim$0.955 and $\sim$0.946
regardless of E$_{vis}$, respectively.
These energy independent features
can provide interesting results
for the existence of a hypothetical ${\nu}_{s}$.
The present proposed ${\bar{\nu}}_{e}$ source can also be useful for neutrino disappearance studies
with future's gigantic LS detectors (LSDs)
such as LENA (Low Energy Neutrino Astronomy) \cite{LENA_0}.
A 50 kt LSD LENA type detector would
have specific features of low energy detection threshold, good energy resolution,
and particle identification with efficient background discrimination
\cite{LENA_a1, LENA_a2, LENA_a3}.
The fiducial volume size for LENA
is set to be
14 m in the radius and
100 m in the height \cite{LENA_a1},
and 3.3 $\times$ 10$^{33}$ target protons
in the volume (44 kt) is assumed \cite{LENA_0}.
The $^{8}$Li generator is placed
at this LS detector
as shown in Fig.~\ref{figIBD_2}(a).
The expected event rates are obtained within
the cylinder shape of LS LENA type detectors,
whose values with
P$_{3}$, P$_{3+1}$ and P$_{3+2}$
and their ratios
are plotted in Fig.~\ref{figIBD_2}(b).
The ratios of R$_{3+1}$/R$_{3}$
are nearly constant with respect to E$_{vis}$.
At the energies of E$_{vis}$ $>$ 6 MeV, however,
the ratio of R$_{3+2}$/R$_{3}$
decreases as
E$_{vis}$ increases.
The comparison between the ratios of
R$_{3+2}$/R$_{3}$ and R$_{3+1}$/R$_{3}$
in this E$_{vis}$ region can give a meaningful signal for
distinguishing
the 3+1 or 3+2 sterile neutrino scenarios.
These characteristics are unique features
of the present work due to the relatively higher energy of
${\bar{\nu}}_{e}$ (E$_{{\bar{\nu}}_{e}}$ $<$ 13 MeV)
and the compact ${\bar{\nu}}_{e}$ source.\footnote{\protect
It is difficult to see the difference in Fig.~\ref{figIBD_1}(b)
by the $^{144}$Ce-$^{144}$Pr antineutrino generators \cite{Ce144s2, Ce144s3}
due to the low energy of
${\bar{\nu}}_{e}$ (E$_{{\bar{\nu}}_{e}}$ $<$ 3 MeV).
}
%

\begin{table}
\caption{The key parameters used in this work.}
\begin{tabular}{c|ccc}
\hline\hline
  Neutron source               & & $^{252}$Cf  &\\
  Neutron intensity            & & 2.34 $\times$ 10$^{12}$ n/s/g &\\
  Neutrino production target   & & $^{7}$Li (99.99\% enhanced) surrounded by graphite &\\
  Run period                   & & 5 years &\\
  ${\bar{\nu}}_{e}$ / neutron  & & 0.256 &\\
  ${\bar{\nu}}_{e}$ flux       & & 6 $\times$ 10$^{11}$ ${\bar{\nu}}_{e}$/s/g &\\
  Neutrino energy cut          & & 7 MeV (E$_{vis}$= 6.22 MeV) &\\ \hline\hline
	Detectors   		       & hemisphere  type & cylinder type &\\ \hline
  Fiducial mass                & 10 kt       & 44 kt        &\\
  IBD event total              & 9750        & 4960        &\\
\hline
\end{tabular}
\label{table_1}
\end{table}

\begin{figure}[tbp]
\centering
\includegraphics[scale=0.4]{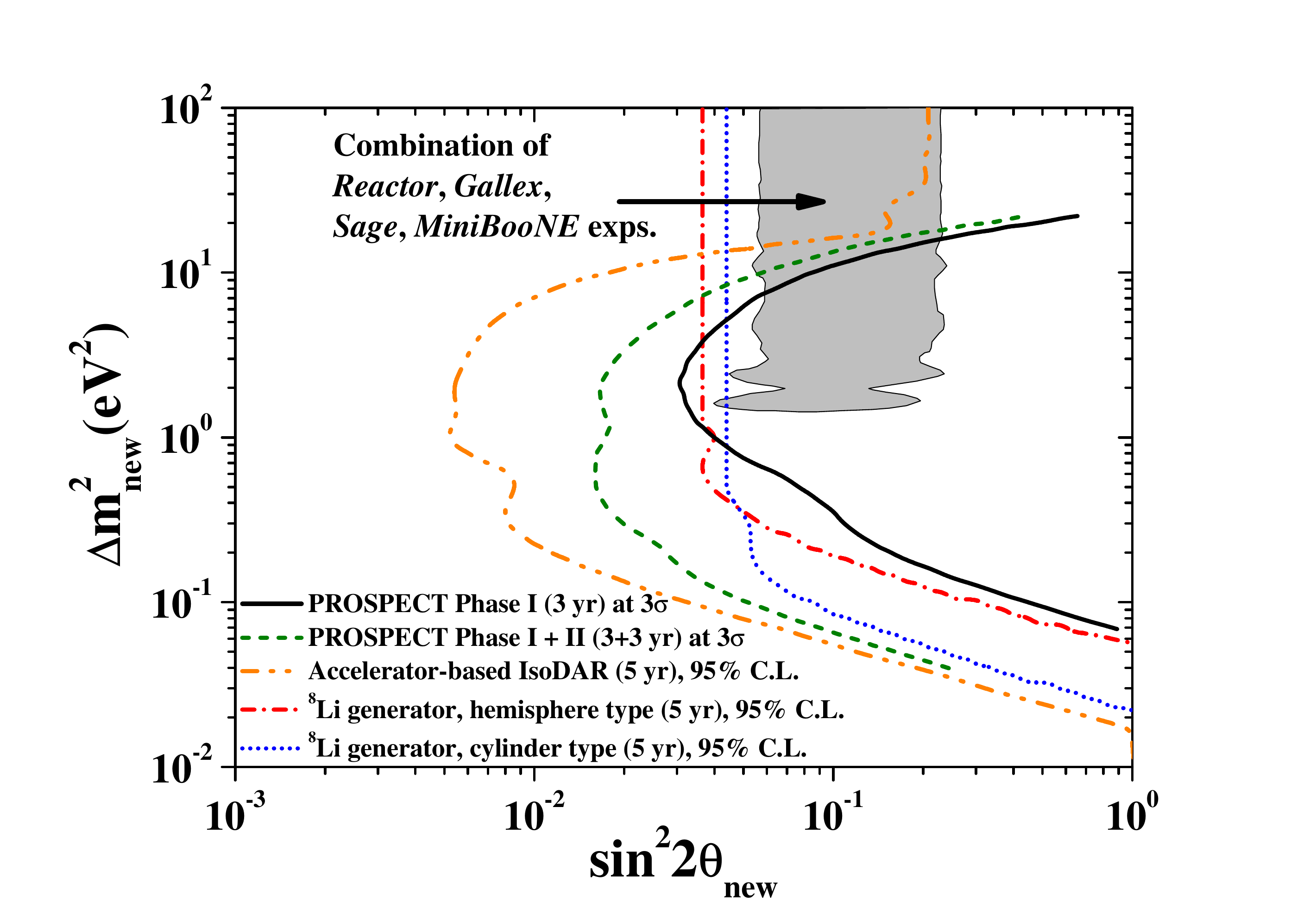}
\caption{(Color online)
95\% C.L. sensitivity of the hemisphere and the cylinder type detector
with $^{8}$Li generator proposed in this work.
The gray area is an allowed region in the parameter space
from the combination of reactor neutrino experiments,
Gallex and Sage calibration sources experiments, and MiniBoone~\cite{reacAnt}.
Sensitivities of PROSPECT (for Phase I and Phase II) \cite{prospect_1} 
and an accelerator-based IsoDAR by using $^{8}$Li source \cite{sterileNu1} 
are also plotted.
}
\label{sen95}
\end{figure}

%
The 95\% C.L. sensitivities of the hemisphere and the cylinder type detectors
with the $^{8}$Li generator are
obtained by following the Eq.(3) of Ref.\cite{Ce144s1}
assuming 3\% systematic uncertainty and
5\% normalization uncertainty.\footnote{\protect
The $^{8}$Li decays mainly to the broad 3.03 MeV, 2$^{+}$ level (first-excited state) of $^{8}$Be,
which then breaks up into two $\alpha$ particles ~\cite{ELevel}.
Experimental $^{8}$Li yields can be obtained by measuring $\sim$ 3.03 MeV of
typical gamma-rays peaks from the $^{8}$Be* in the $^{8}$Li generator
where the peak does not appear for the generator without Li.}
The key parameters adopted for the sensitivity test are tabulated in Table \ref{table_1}, and
whole results are plotted in Fig.~\ref{sen95}, 
where we compared other sensitivities reported from Refs. \cite{reacAnt,prospect_1,sterileNu1}. 
For PROSPECT proposal \cite{prospect_1}, 
the High Flux Isotope Reactor (HFIR) \cite{hfir} with a power of 85 MW and 
two neutrino detectors, AD-I and AD-II, were considered. 
And fiducialized target mass of 1.48 t ($\sim$ 7 t) and 
baseline range of 7 $\sim$ 12 m (15 $\sim$ 19 m) 
were assumed for AD-I (AD-II). 
The sensitivity of an accelerator-based IsoDAR by using $^{8}$Li source \cite{sterileNu1} 
was obtained by considering 
a proton accelerator with a power of 600 kW, 
a fiducial target mass of 897 t 
and 16 m distance between the target face and the center of the KamLAND detector \cite{Kam}. 
Here we exploited the following 2-$\nu$ oscillation survival probability
\begin{eqnarray}
P
= 1 - {\rm{sin}}^{2}(2 \theta_{\rm{new}}){\rm{sin}}^{2}(1.27 \frac{\Delta m^{2}_{\rm{new}}[eV^{2}]L[m]}{E[MeV]})~,
\label{eq:P_new}
\end{eqnarray}
where $\theta_{\rm{new}}$ and $\Delta m^{2}_{\rm{new}}$
are the new oscillation parameters.
When a five-years run with 1.5 g of $^{252}$Cf
is considered, the total expected events
considering the effect of a half-life of $^{252}$Cf
and the neutrino energy cut of 7 MeV
with P$_{3}$
are 9750 and 4960 for the hemisphere and the cylinder type detectors, respectively.
The results in Fig.~\ref{sen95}
clearly show that the proposed neutrino source can cover the region in parameter space from
the reactor anomalies for the two LS detectors,
and thus our scheme can be effectively used for testing the ${\nu}_{s}$ hypothesis
\cite{reacAnt}.



In summary, ${\bar{\nu}}_{e}$ source by the $^8$Li generator with the neutron emitter $^{252}$Cf is compact
so that neutrino detectors can be placed within a few meters
from this neutrino source
with E$_{{\bar{\nu}}_{e}}$ $<$ 13 MeV.
Therefore,
it is an efficient neutrino source
for the study of 1 eV mass scale ${\nu}_{s}$
as well as other neutrino oscillation study.
Two different types of experiments are considered.
One is the hemisphere type detector,
where the $^{8}$Li generator is placed at the center
of the detector.
Another is to place the generator at
the cylinder type detector.
The shapes of event rates
in Figs.~\ref{figIBD_1}(b) and~\ref{figIBD_2}(b)
for the hemisphere and the cylinder type detectors
can give effective chances to search for
the existence of ${\nu}_{s}$
and to test the 3+1 or 3+2 sterile neutrino scenarios.
If we can measure
higher than or equal to
5\% deviation from the
expected events, in the case that the
P$_{3}$ model is true,
we can conclude which of the P$_{3}$, P$_{3+1}$, and P$_{3+2}$
models is the most appropriate scenario.
Together with
the reactor anomaly,
our electron-antineutrino source can
also
be used for
a precise test of the weak mixing angle
at $\Delta m^{2}$ $\sim$ 1 eV$^{2}$ order scale
relevant to the sterile neutrino as shown in the sensitivity test results in Fig.~\ref{sen95}.

\section*{Acknowledgments}
The work of J. W. Shin is supported by
the National Research Foundation of Korea \ (Grant No. NRF-2015R1C1A1A01054083),
the work of M.-K. Cheoun is supported by
the National Research Foundation of Korea \ (Grant No. NRF-2014R1A2A2A05003548 and NRF-2015K2A9A1A06046598).

\section*{References}
\bibliography{mybibfile}

\end{document}